\documentstyle[preprint,aps]{revtex}
\topmargin - 40 pt
\begin{document}               
\draft
\title{Production of bound triplet $\mu^+ \mu^- $ system 
in collisions of electrons with atoms}
\author{N.~Arteaga-Romero$^{(a)}$, C.~Carimalo$^{(b)}$, and 
V.G.~Serbo$^{(b,c)}$}
\date{\today}

\address{$^{(a)}$Universit\'e Paris VI,
4 place Jussieu, Paris Cedex 05 France}

\address{$^{(b)}$Laboratoire de Physique Nucl\'eaire et de Hautes Energies, 
IN2P3-CNRS\\ 
Universit\'e Paris VI \& Paris VII,
4 place Jussieu, Tour 33, F-75252, Paris Cedex 05 France\\
e-mail~: carimalo@in2p3.fr}

\address{$^{(c)}$Novosibirsk State University, Physics Department, 
Novosibirsk, 630090 Russia\\
e-mail~: serbo@math.nsc.ru}
\maketitle
\vskip 10 true cm

\abstract{
This paper deals with the production of 
orthodimuonium (OM) ($\mu^+ \mu^- $ atom in triplet state)
in collisions of high-energy
electrons with nuclei or atoms. This reaction was previously studied 
by Holvik and Olsen [Phys. Rev. D {\bf 35}, 2124 (1987)] 
on the basis of a bremsstrahlung mechanism where OM is produced by 
only one virtual photon. In the present paper we consider a competing 
three-photon
mechanism where the production of OM results from the collision of a  
photon generated by the electron with two photons emitted by the nucleus. 
We derive the corresponding energy spectrum 
and production rate of OM and show 
that the three-photon mechanism is the dominant one for heavy atom target.
}

\section{Introduction}

It is known that $\mu^+ \mu^- $ atoms exist in two spin states: 
paradimuonium (PM, singlet state $n\,^1S_0$) with lifetime
in the ground state $\tau_0 = 
0.6\cdot 10^{-12}$ s and orthodimuonium (OM, triplet state $n\,^3S_1$) 
with lifetime in the ground state $\tau_1 = 1.8\cdot 
10^{-12}$ s. The dominant
decay processes are
\begin{equation}
{\rm PM} \to \gamma \gamma, \;\; {\rm OM} \to e^+ e^-\,.
\label{1}
\end{equation} 

The main physical motivation to study dimuonium production
lies in the fact that dimuonium is one of the simplest hydrogenlike atom, 
that is very convenient for testing fundamental laws. Up to now
there is a lot of theoretical predictions on dimuonium properties 
(for a review see, for example, proceedings~\cite{P}) but dimuonium has 
not been observed yet.\\

A promising method to create dimuonium --- its
production at relativistic heavy ion colliders ---
has been recently considered in Ref.~\cite{GJKKSS}. Another
attractive possibility is the electroproduction of dimuonium
on atoms. The PM electroproduction cross section was calculated in 
Ref.~\cite{MSS} in the main logarithmic approximation, and discussed
in more detail in Ref.~\cite{HO}. In these two papers 
a two-photon mechanism was considered where the incident electron and the 
nucleus emit each virtual photons which then collide to produce a C-even 
PM state.\\

In Ref.~\cite{HO} the electroproduction of C-odd OM state on atoms was 
assumed to proceed from the brems\-strah\-lung mechanism of Fig. 1.
The obtained spectrum of OM production on nucleus of
charge $Ze$ is    
\begin{equation}
{d\sigma_{\rm br}(x)\over dx} = \sigma_0\, f_{\rm br}(x), \;\;
x= {E_{\mu\mu} \over E_e}\,, \;\;
\sigma_0 = {\zeta (3)\over 4}\, {\alpha^5 (Z\alpha)^2
\over m_\mu^2}\,,
\label{2}
\end{equation} 
$$
f_{\rm br}(x) = {x(1-x)\over (1-x+\varepsilon)^2}\,
\left(1-x+{1\over 3} x^2\right)\, 
\left[ \ln{(1-x)^2E_e^2 \over (1-x+\varepsilon) m_\mu^2}\, -\,
1 \right]\,, \;\; \varepsilon = {m_e^2 \over 4 m_\mu^2}
$$
where $E_e$ and $E_{\mu\mu}$ are the energies of electron and OM 
respectively, and $m_e$ and $m_\mu$ are the electron mass and the muon mass 
respectively; $\alpha \approx 
1/137$ and $\zeta (3) = 1.202...$ . The obtained cross section 
has the form
\begin{equation}
\sigma_{\rm br} = 7.16\,\sigma_0\, (L - 2.73)\,,\;\;
L= \ln{E_e\over 2m_\mu}
\label{3}
\end{equation} 
(note that this cross section is positive only for $E_e >
3.2$ GeV).\\

However, the analysis of Ref.~\cite{HO} is incomplete since it did not take 
into account the important three-photon mechanism depicted in Fig.~2. Our aim 
in the present paper is to improve this previous analysis by estimating the 
production rate provided by the latter mechanism and by pointing out 
some of its main features. As we shall show, 
the three-photon process competes with the bremsstrahlung one 
and even predominates in the case of high-energy 
electron scattering by heavy atom target.

\section{Calculation of three-photon orthodimuonium production}

At high electron energy ($E_e \gg m_\mu$) the cross section $d\sigma_{3\gamma}$
corresponding to the diagram of Fig.~2 can be calculated 
using the equivalent-photon approximation. In this approximation 
the cross section $d\sigma_{3\gamma}$ is expressed as the product of
the number of equivalent photons $dn_{\gamma}(x,Q^2)$ generated by the
electron, by the cross section $d\sigma_{\gamma Z}$ for the real  
photoproduction of OM on the nucleus
$$
d\sigma_{3\gamma} = dn_\gamma \, \sigma_{\gamma Z}\,,\;\;
dn_\gamma (x,Q^2)= {\alpha \over \pi}\, {dx \over x}\,
{dQ^2 \over Q^2}\, \left[ 1-x+ {1\over 2} x^2 -
(1-x) \, {Q^2_{\min} \over Q^2} \right]\,,
$$
\begin{equation}
Q^2 = -q^2\,, \;\; Q^2_{\min} = {x^2 m^2_e\over 1-x}\,.
\label{4}
\end{equation} 
After integration of $dn_\gamma (x, Q^2)$ over $Q^2$ from $Q^2_{\min}$ 
up to $Q^2 \sim m^2_\mu$ we obtain the energy spectrum of equivalent
photons 
\begin{equation}
{dn_\gamma (x) \over dx} = {\alpha \over \pi}\, {1\over x}\, 
\left[ \left(1-x+{1\over 2} x^2 \right)\,
\ln{(1-x)m_\mu^2 \over x^2 m_e^2} \,-\,1+x\, \right]\,.
\label{5}
\end{equation} 
The accuracy of this expression is a logarithmic one, i.e.
the omitted items are of the order of $1/l \sim 1/15$ where 
$l = \ln{[(1-x)m^2_\mu /(x^2 m_e^2)]}$.

The photoproduction cross section $\sigma_{\gamma Z}$ can be 
found in Ref.~\cite{GJKKSS}. At high photon energy ($xE_e\gg m_\mu$), 
the corresponding amplitude for photoproduction of OM in $n\,^3S_1$ state 
on nucleus of mass $M_Z$ takes the form
\begin{equation}
M_{\gamma Z} = 4i \alpha^{4} Z^2 {2 x E_e M_Z \over n^{1/3}} \int \,
{F({\bf k}^2_{1\perp})\,F({\bf k}^2_{2\perp})\over
{\bf k}^2_{1\perp} \; {\bf k}^2_{2\perp}}\, \; {\bf e}_\gamma\,.
{\bf e}^*_{\rm OM} \; \times 
\label{5a}
\end{equation}
$$
\left[
{4m_\mu^2 \over 4m_\mu^2 + {\bf p}_\perp^2} -
{4m_\mu^2 \over 4m_\mu^2 + ({\bf p}_\perp -2{\bf k}_{1\perp})^2}
\right]\; 
\delta(
{\bf k}_{1\perp}+{\bf k}_{2\perp} - {\bf p}_\perp)\;
d^2 {\bf k}_{1\perp}\,d^2 {\bf k}_{2\perp}
$$
where $F({\bf k}^2_{i\perp})$ is the nucleus form factor, 
${\bf p}$ is the momentum of OM, ${\bf e}_\gamma$ and 
${\bf e}_{\rm OM}$ are the polarization vectors for the equivalent
photon and the final OM respectively. 
Because of the rapid decrease of the nuclear form factor at large 
transverse momenta, the main contribution to the amplitude
comes from values of tranverse momenta such that 
\begin{equation}
{\bf k}_{i\perp}^2 \stackrel{<}{\sim} {1\over 
\langle r^2 \rangle} 
\label{6}
\end{equation} 
where $\langle r^2 \rangle $ is the mean square radius of the 
charge distribution of the nucleus. We use below the parameter
$\Lambda$ defined as
\begin{equation}
{1\over 6}\, \langle r^2 \rangle = {1\over \Lambda^2}\,,
\;\; \Lambda = {405 \over A^{1/3}}\;\; {\rm MeV}
\label{7}
\end{equation}
where $A$ is the atomic mass number, $\Lambda = 70$ MeV for
Pb and $\Lambda = 120$ MeV for Ca. The cross section 
$\sigma_{\gamma Z}$ is found to be a constant at high photon energy
$xE_e \gg m_\mu$. It was calculated in Ref.~\cite{GJKKSS}
by numerical integration of Eq.~(\ref{5a}), 
using a realistic nuclear form factor. After summing up over all $n\,^3S_1$ 
states of OM, the final result is
\begin{equation}
\sigma_{\gamma Z} = 4\pi \alpha \sigma_0\, B \,
\left({Z\Lambda \over m_\mu}\right)^2, \;\;\; B=0.85\,.
\label{8}
\end{equation}

As a result, the shape of the energy spectrum of OM is  
just that of the equivalent photon spectrum : 
\begin{equation}
{d\sigma_{3\gamma}(x)\over dx} = \sigma_0 \,
f_{3\gamma }(x)\,, 
\label{9}
\end{equation} 
$$
f_{3\gamma }(x)=\left({Z\alpha \Lambda \over m_\mu}\right)^2 
\; {4B\over x}\,
\left[ \left(1-x+{1\over 2} x^2 \right)\,
\ln{(1-x)m_\mu^2 \over x^2 m_e^2} \,-\,1+x\, \right]\,.
$$ 
The total cross section is obtained after integration over
$x$ (from $x_{\min} =2m_\mu /E_e$ up to $x_{\max}=1$)
$$
\sigma_{3\gamma }= 4B\, \sigma_0\,
\left( {Z \alpha \Lambda \over m_\mu}\right)^2\,
\left[ L^2 + \left(L- {3\over 4} \right)\,
\ln{m_{\mu}^2\over m_e^2}\,-\,L - {\pi^2\over 6}-
{1\over 8} \right]=
$$
\begin{equation}
=3.4\, \left({Z \alpha \Lambda \over m_\mu}\right)^2\,\sigma_0\,
\left[ L^2 +9.7\, (L-1)\right]\,.
\label{10}
\end{equation} 

\section{Results and Conclusions}

1. Let us compare the energy and angular distributions corresponding to  
the bremsstrahlung and to the three-photon production of OM respectively. The 
energy spectra are given by Eqs.~(\ref{2}), (\ref{9}) or by the 
curves for $f_{\rm br}(x)$ and $f_{3\gamma}(x)$ presented in Fig.~3.
The bremsstrahlung function $f_{\rm br}(x)$ has a peak at
large $x$ ($x\approx 1$); it depends on the electron energy
but not on the target properties. On the other hand,
the three-photon function $f_{3\gamma}(x)$ has a peak at small
$x$; it does not depend on the electron energy but it 
strongly depends on the type of nucleus.

The respective angular distributions are also different. The typical 
emission angle of OM in bremsstrahlung 
production was estimated in Ref.~\cite{HO} as
\begin{equation}
\theta_{\rm br} \sim {m_\mu\over E_e} 
\label{11}
\end{equation} 
The analogous typical angle in three-photon production is of the order of
(see Eq.~(\ref{5a}))
\begin{equation}
\theta_{3\gamma} \sim {\Lambda \over xE_e}\,
\label{12}
\end{equation} 
and is thus much larger than $\theta_{\rm br}$ in the range $x \ll 1$, 
where the three-photon process dominates.\\

Taking into account the quite different energy and angular 
distributions of OM in the two production mechanisms, we can 
conclude that the interference between bremsstrahlung and three-photon
productions should be very small.\\

2. The total cross sections for the production 
mechanisms here discussed are given by Eqs.~(\ref{3}), (\ref{10}) and  
are presented as functions of the electron energy in Figs.~4 and 5, 
for Pb target and Ca target 
respectively. It is
clearly seen that the bremsstrahlung mechanism is the most important 
for electroproduction on light nuclei while the three-photon 
mechanism is dominant for electroproduction on heavy nuclei.\\

The knowledge of the total cross section gives us a possibility
to estimate the production rate. A detailed procedure to obtain such a 
number can be 
found in Ref.~\cite{HO} where the rate for 
bremsstrahlung production of OM was estimated. For example,
this rate in the case of Pb target is about 3 orthodimuonia per minute
for an electron energy $E_e = 10$ GeV and electron current
of $1$ mA. According to Fig.~4 the rate provided by the three-photon 
orthodimuonium production for the same example is 2.5 times larger 
--- about 7 orthodimuonia 
per minute.

3. Thorought this paper we have considered the electroproduction
of OM on nuclei. Let us briefly discuss the electroproduction 
of OM on atoms where a possible atomic screening has to be 
taken into account. It is known that the atomic screening becomes
important when the minimal momentum transfered to atom $2 m_\mu^2
/E_{\mu \mu} = 2 m_\mu^2 / (xE_e)$ becomes comparable with
the typical atomic momenta $\sim m_e \alpha Z^{1/3}$. 
In other words, the atomic screening should be taken into 
account when the electron energy $E_e$ becomes of the order 
of or larger than the characteristic energy
\begin{equation}
{2m_\mu^2 \over m_e \alpha Z^{1/3}} = {6000 \over
Z^{1/3}}\;\; {\rm GeV}\,.
\label{13}
\end{equation}

4. In the present paper we have calculated three-photon production
of OM due to diagram of Fig.~2 with two photons being exchanged with the 
nucleus. But
the C-odd $\mu^+ \mu^-$ bound system can also be produced in collision
of a photon generated by the electron, with $4,\; 6,\;
8, ...$ photons exchanged with the target. At first sight, this leads to
an additional factor $(Z\alpha)^{n-2}$ for $n$ exchanged photons. 
This is just the case for electroproduction of orthopositronium
on nuclei. Since for heavy nuclei the parameter $Z\alpha$  is not 
small (for example, $Z\alpha = 0.6$ for Pb), the whole series in 
$(Z\alpha)^2$ has to be summed. According to Ref.~\cite{GKSST}, the 
high-order $(Z\alpha)^2$ effects decrease the orthopositronium 
production cross section by about 40~\% for the Pb target.     
However, for OM production we should take into account 
the restriction of the transverse
momenta $k_{1\perp},\; k_{2\perp}, ..., \; k_{n\perp}$ due to
nuclear form factors on the level given by Eq.~(\ref{6}). As a
result, the effective parameter of the perturbation theory
is not $(Z\alpha)^2$ but
\begin{equation}
{(Z\alpha)^2\over \langle r^2 \rangle\, m_\mu^2 } < 0.03
\label{14}
\end{equation} 
and, therefore, we can restrict ourselves to three-photon production
only.

\section*{Acknowledgments}

The authors wish to thank Prof.~I. F. Ginzburg for useful 
discussions. \\
V.G. Serbo is grateful to University Paris VI for a grant that allowed 
his stay in Paris where this work was completed, and to the 
Laboratoire de Physique Nucl\'eaire et de Hautes Energies (LPNHE) for 
warm hospitality. 

\begin{figure}[htbp]
  \begin{center}
    \leavevmode
    \caption{Bremsstrahlung production of orthodimuonium on nucleus}
    \label{fig:1}
  \end{center}
\end{figure}
\begin{figure}[htbp]
  \begin{center}
    \leavevmode
    \caption{Three-photon production of orthodimuonium on nucleus}
    \label{fig:2}
  \end{center}
\end{figure}
\begin{figure}[htbp]
  \begin{center}
    \leavevmode
    \caption{The orthodimuonium spectra $(1/\sigma_0)\, (d\sigma /dx)$
with $\sigma_0 = 0.3\, \alpha^5 (Z\alpha)^2 /m_\mu^2$ for
three-photon production on Pb (solid line) and Ca (dashed line);
for bremsstrahlung production at $E_e=100$ GeV (dot-dashed line),
at $E_e=10$ GeV (dot-long-dashed line) and at $E_e=5$ GeV (dotted line)}
    \label{fig:3}
  \end{center}
\end{figure}
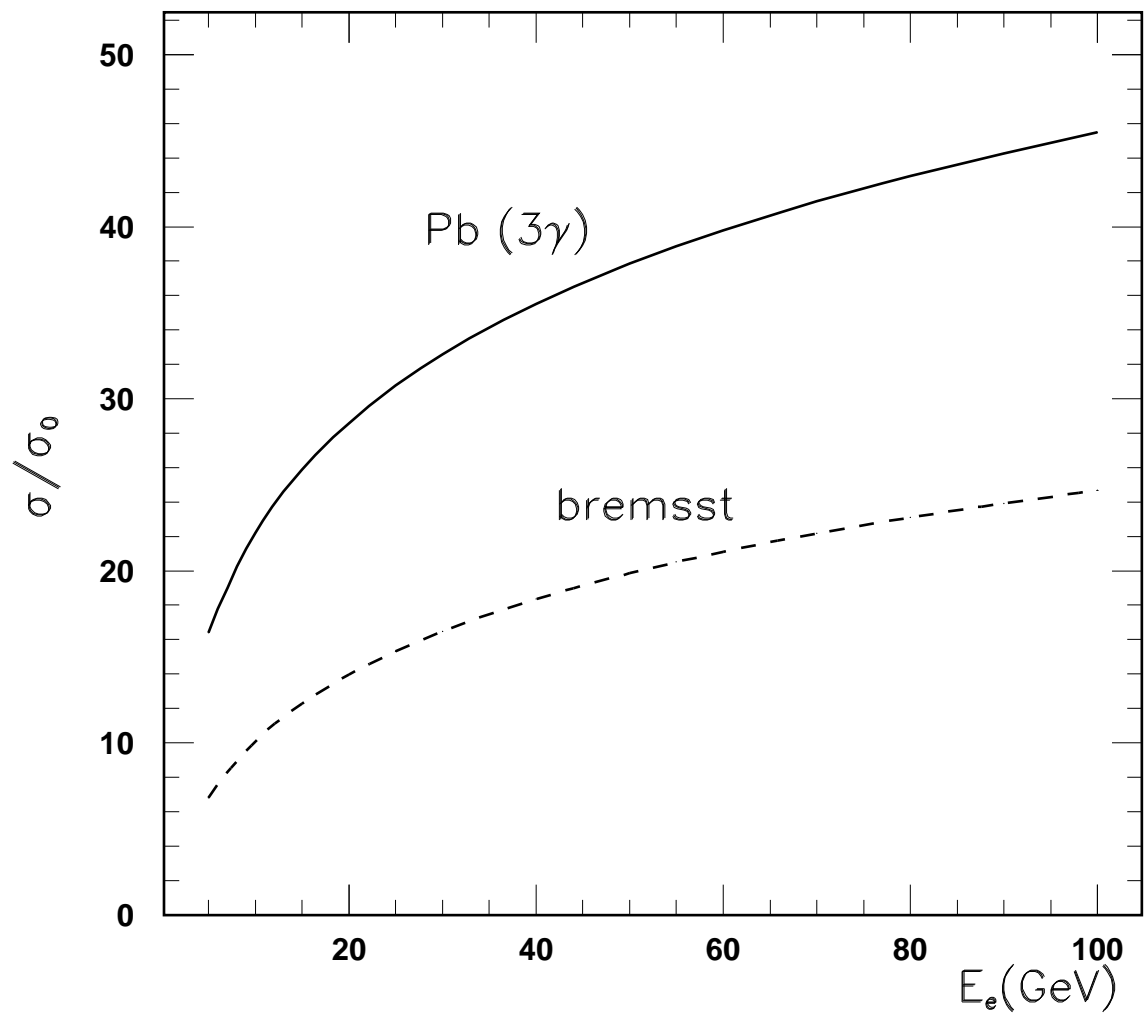
\begin{figure}[htbp]
  \begin{center}
    \leavevmode
    \caption{The total cross section for three-photon (solid line) and
bremsstrahlung (dashed line) production of orthodimuonium on Pb}
    \label{fig:4}
  \end{center}
\end{figure}
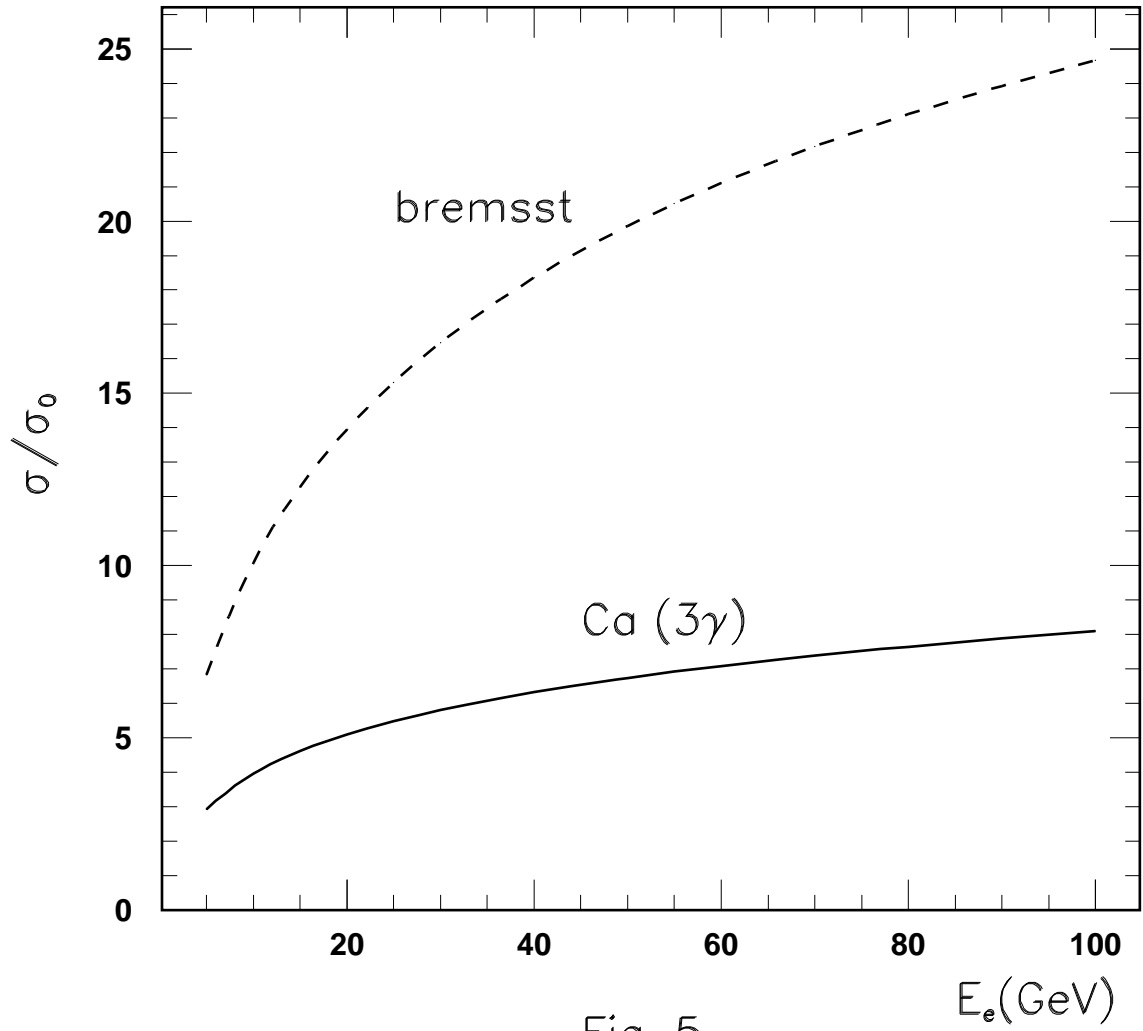
\begin{figure}[htbp]
  \begin{center}
    \leavevmode
    \caption{The same as in Fig.~4 but for Ca target}
    \label{fig:5}
  \end{center}
\end{figure}







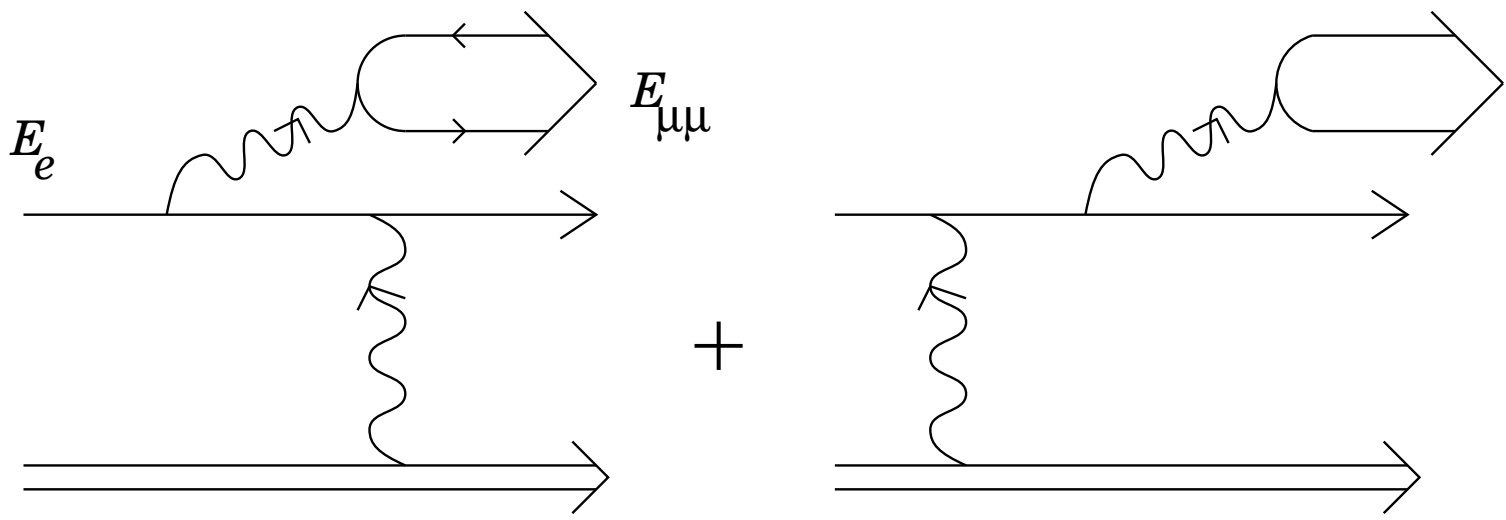
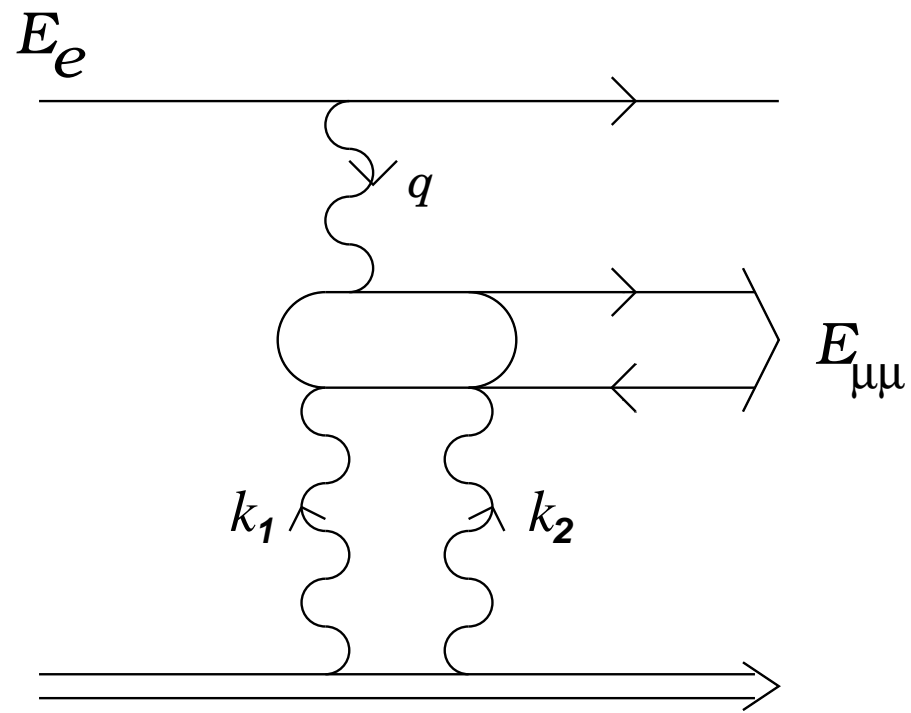
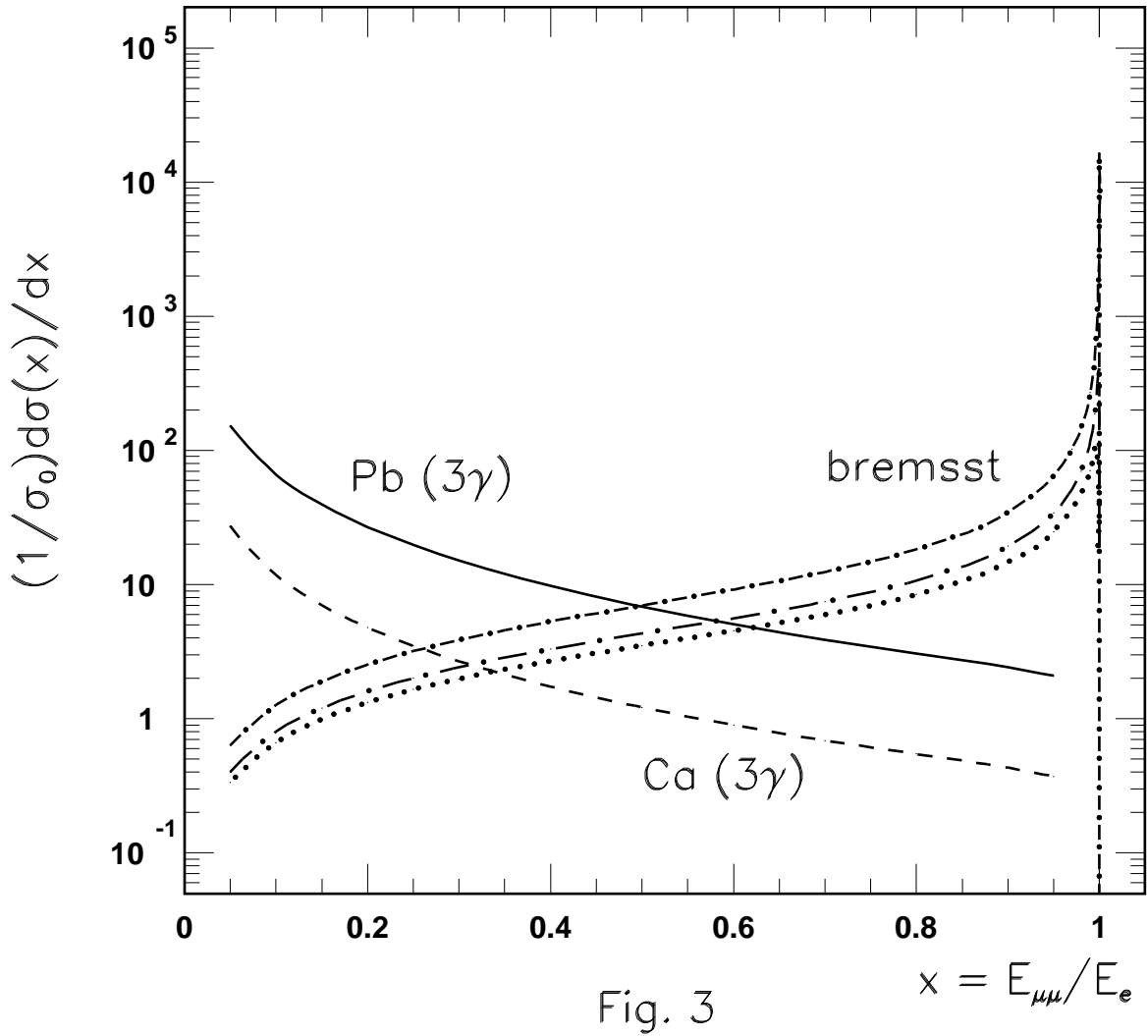
\end{document}